# Female ICT participation in South - Eastern Nigerian Tertiary Institutions: Inhibiting Factors

**Abstract**

The study examined the participation of female students of South – Eastern Nigerian tertiary institutions in Information and Communication Technologies (ICTs). The study discussed the tending gender divide in ICTs participation, reasons for low female participation in ICT, consequences of not bridging the divide and ways of encouraging female participation in ICT. A structured questionnaire was used to elicit information from respondents. A multi – stage random sampling technique was used in the selection of respondents. One hundred and thirty – six (136) undergraduate female students of tertiary institutions in South - Eastern Nigeria constituted the study sample. Data collected was analysed using descriptive statistics. Findings suggest that high cost of ICT and high level of male dominance, which made females think that ICT is for males were the major reasons for low female participation in ICT. Reducing the cost of Information Technology, and parental involvement in their children's selection choice of study were suggested to encourage female participation in Information and Communication Technologies.

**Chinyere A. Nwajiuba**

Department of Educational Foundations and Administration

Imo State University, Owerri

E-mail: canwajiuba@yahoo.com

**Elochukwu Ukwandu**

Department of Computer Science

Imo State University, Owerri

E- mail: eloukwandu@msn.com

**Introduction**

Information and Communication Technologies (ICTs) refer to technologies, which are being used for collecting, storing, editing and passing on information in various forms (SER, 1997 as cited in Olubamise, 2001). ICT creates earning opportunities, improve delivery and access to health and education, facilitate information and increase the transparency, accountability and effectiveness of government, business and non-profit organisations. This again creates an environment for development of a nation like Nigeria. As United Nations development programme (2002) puts it, making ICT an integral part of development cooperation, developing countries and their partners can more efficiently address economic and social divides.

Information and Communication Technologies (ICTs) in recent times have been seen as a rallying point for faster economic and social development of any nation. This is peculiar with nations which information is an essential and valuable commodity that one can buy, sell, store and exchange. It has become obvious that information is what people need to change their world. This is only possible if this information is shared and distributed all around the world enabling all people to access knowledge and benefit from being educated. Without widespread of these information, the Millennium development Goals set by the international community will be much harder to meet.

ICTs enable a knowledge network for people especially the youths in higher institutions. This is because they are the potential contributors to the global market. They can use ICTs to participate in and complete various learning tasks, whether formal or informal (Olubamise, 2007). ICTs can also be used in searching and collecting educational materials, which help to ensure their future success within their future workplaces. This eventually improves the countries economic capital. ICTs are for everyone both males and females. Both should therefore, have to be an equal beneficiary to the technologies, and the products and processes, which emerges from their uses (Vikas, 2003). A personal Computer and Internet facility are the best-known example of the use of ICT in Education. In this paper, ICTs refer to use of personal Computers and Internet facilities.

**Problem Definition**

Over the years, the participation of women in ICTs have been a cause for concern. Female participation in Nigeria is low (Liverpool, 2005). ICT participation is meant to be for all gender for national development and humanity in general. Given that women comprise over half of the world's population, development cannot be possible without their inclusion and support. It is supposed to be designed in relation to gender differences and handicaps for proper representation. This implies a technology that is gender sensitive. In USA recent survey by a Stanford based research firm, Garter Inc, predicts that by 2012, 40% of women now in information technology workforce will move away from technical career paths to pursue more flexible business, functional, research and development careers (Brian, 2007). This has become necessary because of the drive of having a gender divide within a digital divide already existing in developing nations. Preventing this is of utmost importance because women are moulders of the nation for there is a highly acclaimed saying that, "if you educate a woman, you educate a nation". It is an establishment fact that investment in a woman is of value not only to her individually, but to the wider society (GC, Women's Ministry, available at http://wm: gc.adeventist.org/pages/literacy.2html).

This paper is therefore, primarily concerned about answering the following questions; what constitute the hindrances to the low female participation in technologies especially in ICTs? Are technologies designed with men in view?, what are the major consequences of this trend to nation building? What could be done to achieve a level playing ground for both genders in terms of ICTs participation?

**Research Objectives**

The broad objective is to examine the inhibiting factors on ICTs participation among female students in tertiary institutions.

Specifically, the objectives include,

1) To examine the reasons for low ICT participation among female students.
2) To examine the consequences of the low participation to nation building.
3) To examine ways of encouraging female students participation.

**Research Questions**

1) What are the reasons for low ICT participation among female students?
2) What are the consequences of the low participation to nation building?
3) What are the likely ways of encouraging female students participation in ICT?

# Literature Review

## Female Participation in ICT

While increasing statistics are being made available on female use of ICT in developed countries, it is still quite difficult to get reliable statistics in developing countries especially in Nigeria. However, Ighoroje, (undated), Odedra-Straub (undated) and Hafkin, (2001 as cited in Liverpool, 2005) affirm that very few women participate in ICT in Africa and that ICT is still seen as a man's job in Africa. Liverpool, (2005) observed that over the years, there has been a record of low female participation in ICT. The majority of available literature on gender and Information Technology especially in developing countries appears to depict a disturbingly dismal situation as far as female participation in ICT is concerned. Interestingly, some studies show that there are successful cases of ICT adoption and use by women not ignoring the many persisting problem, which still subsist (Liverpool, 2002).

## Reasons for low female participation in ICT

The issue of gender and ICT is a very complex one with many different reasons proposed for the low female participation in ICT. Most of the reasons are borne out of socio- cultural stereotyping of the roles of women in the society. Ighoroje, (undated) observed in his study that women felt that Information Technology was too tedious, strenuous, too dirty, time consuming, masculine, over expose girls to western styled which hamper their chances of marriage since culture does not allow girls to mix freely with their male counterparts and to feel that what men can do, they too can do.

According to Liverpool, (2005), reasons for low female participation also include non ownership of equipment, less time to commit to learning new programs, what to do when there is problem in browsing the internet, low level of interest shown by women towards ICT. Studies have shown that women tend to be more involved in and interested with the communicative aspect of ICT. It has also been observed that whilst their male counterparts are fiddling with computer parts and other technical issues, women are more interested in attending to customers and helping them deal with computer related problems (Liverpool, 2002). Mentoring is another issue related to low participation of women in ICT. There seem to be very few female ICT experts that girls look up to or to desire to be like (Liverpool, 2005).

Lack of encouragement by both schools and families may contribute to low female participation in ICT. With families, gender roles are stereotyped. When girls observe their mothers in professions other than Mathematics and Sciences, this could have a strong, albeit unconscious, effect on their choice(s).

**Consequences**

At the International Telecommunication Union's (ITU) 2003 World Summit on the Information Society (WSIS), it was recognised that, there is a global need "to build a people centred, inclusive and development oriented Information Society". Which will "create, access, utilize and share information and knowledge" and that connectivity is a central enabling agent in building the information society. Countries that fail to enable access by their citizens are deprived of accessibility to the many benefits of basic and advanced communications, including improved healthcare, education and economic opportunities and the increased ability to participate in the political processes (Manner, 2004). Participation in IT could enable individuals to communicate more with families, friends and colleagues, and also increase social cohesion among others.

According to the United Nations Development Program (2002), low female participation in ICT result to lack of exposure to variety of skills that will allow them to be advanced in jobs, become underpowered and advanced. This is because ICT serve as a tool for enhancing women capabilities in information dissemination (Liverpool, 2002). Low female participation could also result to low representation of female, amongst ICT university professors. There is considerably fewer female than male professors worldwide (Liverpool, 2005). In full computer Science professors are female (Burger, 1999). Low female participation in ICT, could take them away from public life, for instance, in obtaining political positions (http:// www.un.org/women watch/daw/).

**Research Methods**

The research adopts the descriptive survey design. Imo State was the study area. It is one of the thirty-six states in Nigeria. In urban and some part of rural areas, there exist Cyber Cafés and ICT centres. In the higher institutions, there are Cyber Cafés for educational purposes. Imo State is made up of five public tertiary institutions namely, Imo State University, Owerri, Federal University of Technology, Owerri, Alvan Ikoku

College of Education, Owerri, Federal Polytechnic, Nekede, and College of Agriculture, Umuagwo.

**Sample and Sampling Technique**

A multi stage random sampling technique was used in the selection of respondents. Three tertiary institutions were randomly selected for the study. They include, Imo State University, Owerri, Federal University of Technology, Owerri, and Alvan Ikoku College of Education, Owerri. Female undergraduate students constituted the population of the study. Undergraduate female students that cut across different departments and levels were randomly selected from each tertiary institution. A total number of 136 undergraduate students constituted the sample for the study.

**Data Collection**

A structured questionnaire was used in the collection of data. The questionnaire addressed (a) reasons for low female participation in ICT, and (b) suggested ways of encouraging female participation in ICT. The questionnaire is a checklist type and students were allowed to tick off the response they considered most appropriate. They were also given the opportunity of making their views heard. All questionnaire were filled out and all returned afterwards.

Before the questionnaire was given out to the subjects, a test, re-test was done to establish its reliability. A reliability coefficient of a test retest result shows an r' value of 0.88 using the Pearson product moment coefficient.

Data collected was analysed using descriptive statistics.

## Results

**Reasons for low Female students participation in ICT.**

29.4% of the respondents indicated that the reason for low female ICT participation was that ICT is costly, 20.6% indicated that they think that ICT is mostly for men, 14.7% indicated that they can't see female participating in ICT and 14.7% also indicated that it is time consuming (table 1).

**Suggested ways of encouraging participation of females in ICT.**

Results show that 50% of the respondents indicated that reducing the high cost of technology would help encourage female participation, 23.5% suggested that parental involvement in selecting course for their children's study would also help encourage them, 14.7% indicated that breaking of IT into smaller units that are less strenuous and time consuming would encourage these female students, while 11.8% suggested that school authorities should motivate the female students by bestowing awards to the best female students in ICT.

## Discussions

**Reasons for low female participation in ICT**

Results suggest that the major reason for low female participation in ICT is the high cost of involving in it. This implies that to own a personal computer and to use the

Internet facilities are expensive. This could be attributed to the fact that most of these students are from low socio economic status and therefore cannot afford personal computers or even visit Cyber Cafés. Even when the money is available, priority is given to the males because they are believed to be more likely to use their experiences from the technology (Odedra - straub, undated). Liverpool, (2005) also observed that lack of ownership of personal computers is one of the reasons for low female participation in ICT. This is probably due to high cost of computers. Manner, (2004) is of the opinion that ICT should be made women – friendly in terms of cost. Reducing the cost of ICT is the first step to encourage female participation in ICT. This would help encourage the ownership of personal computers, the use of Cyber Cafés and minimise gender disparity in the use of ICT.

Result also indicates that most female students think that ICT is meant for males. This could be attributed to the fact that more male students are always seen using Internet facilities and personal computers than the female students. This view is also supported by Ighoroge, (undated), who observed that the reason for low female participation in ICT was that it looked masculine.

**Ways of encouraging female participation in ICT**

Information and Communication Technology are of paramount importance to the education of students in Nigerian Universities. With knowledge as the crucial input for productive processes within today's economy, the efficiency by which knowledge is gotten and applied determines the quality and economy of students and the nation at large. According to Manner, (2004) the goal of universal access is achieved in a country when ICT services are made available to all citizens. To successfully achieve this goal therefore, a country or state must enable an ICT regime that meets the component of affordability. Reducing the high cost of ICT is of paramount importance. ICT should be affordable to all citizens.

Again, result suggests that parents should be involved when their children especially the female ones are selecting course of study. This is could be explained by the fact that parents being the primary socializing agents could have great influence on their children choices and interests. Fathers and especially mothers who are literate in IT can make positive impart on their female children and serve as mentors. This would help encourage them in participating in ICT.

**Conclusion**

Based on existing literature, it is evident that there is low participation rate of female in ICT worldwide. ICT is indeed a propelling force in the education sector and the global market. Progress of any nation increasingly depends upon the products of educated minds. Women have been proven to be more effective in the experience and do constitute over half of the world's population. Thus, if Nigerian is hoping for long - term development in the recent world, human resources, which obviously include women must be fully harnessed. Based on the reasons for low female ICT participation as found in the study, ICT should be made women – friendly in terms of cost.

**Recommendations**

In addition to the above-mentioned ways of encouraging female ICT participation, the technology being developed must take into account the limited free time available to many women.

There should be the development and promotion of mentoring programs, where women would be trained as instructors, and trainers can serve as role models.

Even as the drive to ensure female access to basic education is strengthened, IT should be integrated into girl's education and women's literacy programs to expose girls to new technology at early stages.

The technology should be gender-friendly, available and enable adaptability. ICT should become "women – friendly" in terms of cost, access applicability in different fields.

Women and girls need to have the same educational opportunity as men and boys. This is because illiteracy among African women constitutes the first obstacle to their use of ICT (Liverpool, 2005).

Relevant areas of interest need to be developed to bring women to ICT tools to enhance participation.

ICT was discussed based available literature. Published studies and public discourse by experts in ICT were used to outline the likely consequences of low participation of female students in ICT and to suggest ways of encouraging female students participation. The above listed objectives were studied.

**Tables 1:**

**Responses on reasons for low ICT participation.**

| Reasons | Frequency | Percentage (%) |
|---|---|---|
| Think it is mostly for men | 28 | 20.6 |
| Peer pressure | 8 | 5.8 |
| Time consuming | 20 | 14.7 |
| Strenuous | 12 | 8.8 |
| Costly | 40 | 29.4 |
| Mentally demanding | - | - |
| Can't see other female in it | 20 | 14.7 |
| Family influence | 8 | 5.8 |
| Total | 136 | 100 |

Source: Field Survey, 2007

Table 2:

Respondents' suggestion on ways of encouraging female ICT participation

| Suggestions | Frequency | Percentage (%) |
|---|---|---|
| Parental involvement in their children's course selection | 32 | 23.5 |
| Breaking of ICT into smaller units that are less strenuous and time consuming | 20 | 14.7 |
| Reducing high cost of technology | 68 | 50 |
| Motivation of female students by school authorities | 16 | 11.8 |
| Total | 136 | 100 |

Source: Field Survey, 2007